\documentclass[useAMS,usenatbib]{mn2e}

\def\cs{IPHASJ211420.03+434136.0} 
\def\n{IPHASXJ211420.0+434136}   
\def\oiii{[O\,{\sc iii}]}
\def\neiii{[Ne\,{\sc iii}]}
\def\ha{H$\alpha$}

\def\hi{H\,{\sc i}}
\def\hb{H$\beta$}
\def\hg{H$\gamma$}
\def\hd{H$\delta$}

\def\nii{[N\,{\sc ii}]}
\def\sii{[S\,{\sc ii}]}
\def\siii{[S\,{\sc iii}]}
\def\ariii{[Ar\,{\sc iii}]}
\def\ariv{[Ar\,{\sc iv}]}
\def\oii{[O\,{\sc ii}]}

\def\hei{He\,{\sc i}}
\def\heii{He\,{\sc ii}}
\def\cii{C\,{\sc ii}}
\def\kms{\relax \ifmmode {\,\rm km\,s}^{-1}\else \,km\,s$^{-1}$\fi}
\def\deg{$^\circ$}
\def\msun{M$_\odot$}
\def\lsun{L$_\odot$}

\def\cb{$c$(H$\beta$)}
\def\Ne{$n_{\mathrm e}$}
\def\Te{$T_{\mathrm e}$}

\usepackage{graphicx}
\usepackage{ulem}

\usepackage[bookmarks=true,bookmarksopen=true,bookmarksopenlevel=10,bookmarksnumbered=true,colorlinks=true,linkcolor=blue,urlcolor=blue]{hyperref}
\title[\n]{The planetary nebula \n\ (Ou5): insights into 
  common-envelope dynamical and chemical evolution}
\author[R.L.M. Corradi et al.]{R.L.M. Corradi$^{1,2}$\thanks{E-mail:
    rcorradi@iac.es}, P. Rodr\'\i guez--Gil$^{1,2}$, D. Jones$^{3,4}$, 
  J. Garc\'\i a--Rojas$^{1,2}$, 
  \newauthor A. Mampaso$^{1,2}$, D. Garc\'\i a--Alvarez$^{1,2,5}$, 
  T. Pursimo$^{6}$, T. Eenm\" ae$^{7,8}$, \newauthor T. Liimets$^{7,8}$, 
and B. Miszalski$^{9,10}$\\
$^{\,1}$Instituto de Astrof{\'\i}sica de Canarias, E-38200 La Laguna, 
Tenerife, Spain\\
$^{\,2}$Departamento de Astrof{\'\i}sica, Universidad de La Laguna, 
E-38206 La Laguna, Tenerife, Spain\\
$^{\,3}$European Southern Observatory, Alonso de Cordova 3107, Casilla 19001, Santiago, Chile\\
$^{\,4}$Universidad de Atacama, Copayapu 485, Copiap\'o, Chile\\
$^{\,5}$Grantecan S.A., Centro de Astrof\'\i sica de La Palma, Cuesta de San
Jos\'e, E-38712 Bre\~na Baja, La Palma, Spain\\
$^{\,6}$Nordic Optical Telescope, Apartado 474, E-38700 Santa Cruz de La Palma, Spain\\
$^{\,7}$Tartu Observatory, Observatooriumi 1, 61602 T\~oravere, Estonia\\
$^{\,8}$Institute of Physics, University of Tartu, Riia 142, 51014 Tartu, Estonia\\
$^{\,9}$South African Astronomical Observatory, PO Box 9, Observatory, 7935, South Africa\\
$^{10}$Southern African Large Telescope Foundation, PO Box 9, Observatory, 7935, South Africa\\
}
\begin{document}

\date{DOI: 10.1093/mnras/stu703 ; BibCode: 2014MNRAS.441.2799C}


\maketitle

\label{firstpage}

\begin{abstract}
While analysing the images of the IPHAS \ha\ survey, we noticed that
the central star of the candidate planetary nebula \n\ (also named
Ou5) was clearly variable. This is generally considered as an
indication of binarity. To confirm it, we performed a photometric
monitoring of the central star, and obtained images and spectra of the
nebula.

The nebular spectrum confirms that \n\ is a planetary nebula of
moderately high excitation. It has a remarkable morphology with two
nested pairs of bipolar lobes, and other unusual features. The light
curve of the central star reveals that it is an eclipsing binary
system with an orbital period of 8.74 hours. It also displays a strong
irradiation effect with an amplitude of 1.5~mag.

The presence of multiple bipolar outflows adds constraints to the
formation of these nebulae, suggesting the occurrence of discrete
ejection events during, or immediately before, the common-envelope
phase.

\n\ also adds evidence to the hypothesis that a significant fraction
of planetary nebulae with close binary central stars have a peculiar
nebular chemistry and a relatively low nebular mass. This may point to
low-mass, low-metallicity progenitors, with additional effects related
to the binary evolution. We also suggest that these objects may be
relevant to understand the abundance discrepancy problem in planetary
nebulae.

\end{abstract}


\maketitle
%

\section{Introduction}\label{Introduction}

Interactions in binary stars are the favoured mechanism to explain the
departure from spherical symmetry observed in a large number of
planetary nebulae \citep[PNe, see e.g.][]{m87,s97,n06}.  However, it
has been traditionally difficult to observationally prove this idea,
because PN central stars are generally faint in the optical, and often
hidden by the emission/absorption of the surrounding gas/dust. In
addition, it is not easy to disentangle the luminosity contribution of
low-mass companions \citep[see e.g.][]{z91,dm13} or interpret radial
velocity measurements \citep{dm07}.  Furthermore, detection of wide
binaries requires high-precision spectroscopy over a long period
\citep{vw14}.

Only in the last few years, starting with the work of \cite{m09a}, the
number of PN binary central stars with known orbital periods has
significantly increased to about forty objects from the initial
  dozen found by \cite{b00}. However, it is important to continue in
the effort, both to increase the statistics, from which general
properties can be derived, and to provide detailed studies of
individual objects, in order to determine which are the relevant
physical processes at work.

\n\ is a candidate planetary nebula identified in the INT/WFC
Photometric \ha\ Survey of the northern Galactic plane
\citep[IPHAS,][]{d05}.  The nebula was independently found by Nicolas
Outters: following the notation used for the other objects discovered
by this French amateur astronomer \citep{acker12,acker14}, the
nebula should also be named Ou5.
The object attracted our attention because its central star is
relatively bright at red wavelengths and presents clear signs of
variability in the three epochs at which IPHAS observed it.  These
characteristics are suggestive of the presence of a binary central
star, where the variability reflects the orbital motion of irradiated
\citep{c11,m11}, gravitationally distorted \citep{s14}, or eclipsing
\citep{m08} stars, or a combination of them \citep{j14}.  We have
therefore gathered photometric and spectroscopic observations of
\n\ that are presented in the following sections.

\section{Observations}\label{observations}

\n\ is located in Cygnus, in a region rich in extended nebulosities
two degrees east of the North America nebula.
The nebula is easily visible in the 2~min \ha+\nii\ exposures of
the IPHAS survey obtained with the 2.5m~Isaac Newton Telescope
(INT). 
Deeper images of \n\ were obtained on November 7, 2013 at the 2.6m
Nordic Optical Telescope (NOT) and the ALFOSC instrument. The filter
central wavelength and full width at half maximum (FWHM) were
6577/180~\AA\ (\ha+\nii) and 5007/30~\AA\ (\oiii). Exposure
time was 30 min in \ha+\nii, and 15 min in \oiii.
The spatial scale of the ALFOSC instrument is $0''.19$~pix$^{-1}$, and
seeing was $0''.95$ FWHM for the $H\alpha$ image, and $0''.75$ for
\oiii.

During the the same night, a 15~min spectrum of the source was
obtained at the 10.4m~Gran Telescopio Canarias (GTC) telescope. The
OSIRIS instrument was used in its longslit mode. Grism R1000B and a
slit width of $1''$ provides a spectral dispersion of 2.1~\AA\ per
(binned $\times$2) pixel, a resolution of 7~\AA, and a spectral
coverage from 3700 to 7800~\AA.  The spatial scale is $0''.254$ per
binned pixel.  The slit was oriented at position angle 311\deg\ and
crossed the central star. A 15~min red spectrum, covering from 5100 to
9200~\AA, was obtained with the same instrumentation on November 15
using grism R1000R and positioning the slit at
P.A. 281.5\deg. Spectral dispersion and resolution were
2.6~\AA~pix$^{-1}$ and 8~\AA, respectively.  Flux calibration was
performed by observing the spectrophotometric standard stars GD48 and
GD191--B2B \citep{oke90}.

\begin{figure}
\centering
\includegraphics[width=\columnwidth]{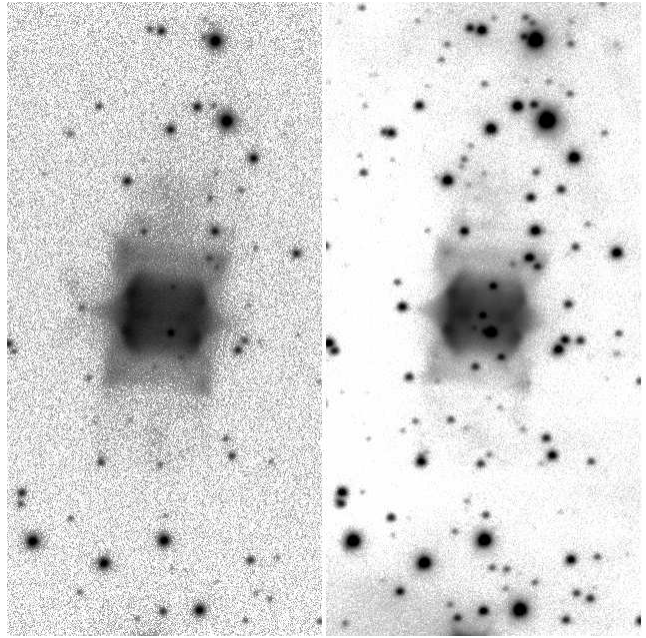}
\caption{NOT \oiii\ (left) and \ha+\nii\ (right) images of \n.  The
  field of view in each panel is 1$\times$2~arcmin$^2$.  North is up,
  east is left.  Greyscale display is in a logarithmic scale. The
  central star is the faint one at the symmetry centre of the nebula
  in the \ha+\nii\ image.}
\label{F-image}
\end{figure}
\begin{figure}
\centering
\includegraphics[width=0.7\columnwidth]{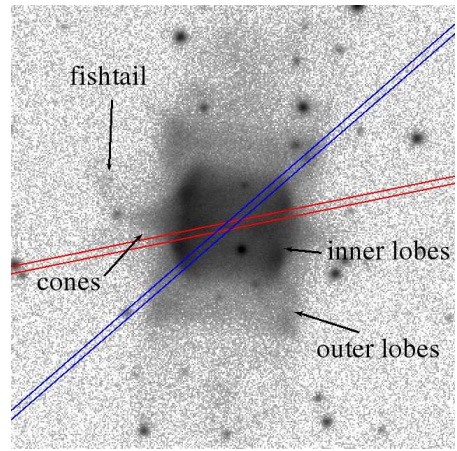}
\caption{Labels of features discussed in the text superimposed on the
  \oiii\ image. The slit position of the GTC spectra (colours
  correspond to the blue and red settings of the spectrograph) is also
  indicated.}
\label{F-sketch}
\end{figure}

Time-resolved photometry of the central star of \n\ in the SDSS
$i$--band was performed during the nights of October 31, and November
14, 23, 24, and 25, 2013, with the 0.8m IAC80 telescope and its
CAMELOT camera. The pixel size is 0$''$.30 and the seeing varied from
1$''$.0 to 1$''$.6. The source was followed during about four hours
every night.  In order to complete the light curve and precisely
determine its periodicity, additional photometry was obtained at other
telescopes. \n\ was observed during a total of nine hours on November
8 and 17 with the 0.6m Zeiss 600 telescope at Tartu Observatory using
an Andor Ikon-L back-illuminated camera and a Bessel I--band filter
from Optec. Here the pixel size is 0$''$.38 and seeing varied from
1$''$.5 to 2$''$.5.  The central star of \n\ was also observed at the
INT on the night of November 28 using its WFC camera (0$''$.33 per
pixel), and with the 4.2m~WHT telescope and ACAM on December 16, both
with a SDSS $i$ filter.  Exposure times varied from 40 sec to 10 min
depending on the telescope and the brightness of the central star.
Magnitudes obtained with the different telescopes were matched to the
IPHAS photometry \citep{bar14} using field stars.

The basic reduction of all data was carried out with standard routines
in IRAF {\it V2.16}\footnote{IRAF is distributed by the National
  Optical Astronomy Observatory, which is operated by the Association
  of Universities for Research in Astronomy (AURA) under cooperative
  agreement with the National Science Foundation.}.

\section{Analysis}
\label{analysis}

\subsection{The nebula}

The NOT images of \n\ are shown in Fig.~\ref{F-image}. The nebula has
a remarkable morphology that is similar in the \oiii\ and
\ha+\nii\ light. It is composed of an inner body with a mild
bipolar morphology and an overall barrel shape of $\sim$14$''$
side. We name it ``inner lobes''. Unusual cone-shaped features
protrude from the waist of the central body along the east-west
direction.  In addition, a second pair of fainter, truncated lobes
extends along the north-south direction: we call them ``outer lobes''.
Along this direction very faint emission extends to as far as 30$''$
from the centre on both sides.  In the \oiii\ image, at the end of the
eastern cone-shaped protrusion, there is evidence of very
faint emission with a fishtail shape. This might be similar to
the structure observed in NGC~6369 \citep{scm92}, or the base
of a third pair of larger and fainter lobes. All these features are
labelled in Fig.~\ref{F-sketch}.  Altogether, \n\ seems to be composed
of two pairs of nested bipolar lobes (and perhaps a third one),
seen edge-on, opened at their poles, with a mild ``equatorial'' waist,
and all sharing the same symmetry axis.

Emission-line fluxes for the nebula were measured in the
GTC spectra.  A precise background subtraction is
somewhat limited by the overlapping and inhomogenous emission of the
extended nebulosities inside which \n\ is projected on the
sky. Looking at the emission line profiles along the slit (before and
after background subtraction), it appears that the most affected line
is the \oii3728 doublet. As this line also falls in a spectral region
where the instrument sensitivity quickly drops, its highly uncertain
flux measurement is not considered in the following analysis.  Other
lines affected by the background emission, although to a lesser
extent, are the \hi\ Balmer lines, \sii\ and \nii.
Tab.~\ref{T-linefluxes} presents line fluxes measured by integrating
the nebular emission covered by the spectrograph slits avoiding the
inner 4 arcsec from the central star. All line fluxes are normalised
to \hb=100 (for the red spectrum, first lines were normalised to \ha,
and then \ha\ was normalised to \hb\ using the blue spectrum).  The
fact that the blue and red GTC spectra do not cover the same portion
of the nebula might explain some differences in the fluxes of
overlapping lines.
The quoted errors on the observed fluxes include both the uncertainty
in the flux of each line (Poissonian, detector and background noise)
and in the determination of the instrumental sensitivity function, but
do not include systematic errors introduced by the subtraction of the
background.

\begin{figure*}
\centering
\includegraphics[width=0.99\hsize]{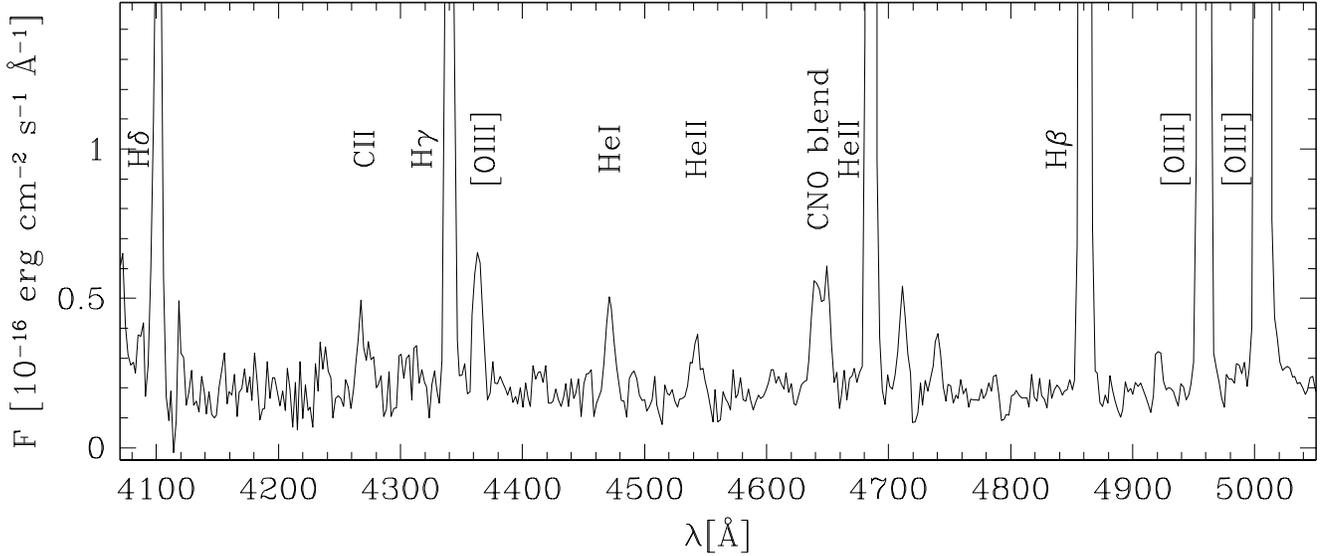}
\caption{Zoom of the blue region of the GTC spectrum of the nebula
  where faint lines discussed in text are located.}
\label{F-spec}
\end{figure*}

\begin{table}
\caption{Observed and dereddened line fluxes in the blue and red GTC
  spectra of the nebula, normalised to H$\beta$=100. In parenthesis
  are percent errors.  The fluxes of \hi\ lines include a
  3.5\%\ contamination by unresolved \heii\ Pickering lines.}
\begin{tabular}{lrrrr}     
\hline\hline\\[-7pt]                
Ident.       & \multicolumn{4}{c}{Flux} \\
             & \multicolumn{2}{|c|}{Blue (P.A.=311\deg)} & \multicolumn{2}{|c|}{Red (P.A.=281.5\deg)} \\
             & Obs.      & Der.  & Obs.      & Der. \\
\hline\\[-5pt]                     
\neiii 3869               &     36.50 (11) &    66.37 (19)    &             &              \\  
\hi\   3889               &      8.94 (14) &    16.06 (21)    &             &              \\  
\neiii+H$\epsilon$        &     21.26 ( 7) &    36.49 (16)    &             &              \\  
\hd\ 4102                 &     19.53 ( 6) &    30.92 (13)    &             &              \\  
\cii\ 4267$^\star$         &      2.36 (50)&     3.38 (51)    &             &              \\
 \hg\ 4340                &     36.14 ( 6) &    49.34 (10)    &             &              \\  
\oiii\ 4363               &      5.62 (10) &     7.56 (13)    &             &              \\  
\hei\ 4471                &      3.82 (15) &     4.81 (18)    &             &              \\
4645 blend                &      9.28 (20) &    10.53 (22)    &             &              \\  
\heii\ 4686               &     63.52 ( 6) &    70.34 ( 6)    &             &              \\  
\ariv\  4711              &      3.74 (15) &     4.07 (18)    &             &              \\  
\ariv\  4740              &      2.08 (30) &     2.23 (33)    &             &              \\  
\hb\  4861                &    100.00 ( 6) &   100.03 ( 6)    &             &              \\  
\oiii\ 4959               &    188.80 ( 3) &   178.75 ( 3)    &             &              \\  
\oiii\ 5007               &    575.28 ( 3) &   530.36 ( 4)    &             &              \\  
\heii\ 5411               &      7.30 (10) &     5.46 (13)    &     6.03(6) &   4.51 (10)  \\  
\hei\  5876               &     31.92 ( 6) &    19.29 (14)    &    30.34(4) &  18.33 (14)  \\  
\nii\ 6548                &     10.86 (10) &     5.06 (23)    &     7.32(6) &   3.42 (21)  \\  
\ha\  6563                &    655.71 ( 3) &   303.99 (20)    &   655.65(3) & 304.00 (20)  \\  
\nii\ 6583                &     39.61 ( 6) &    18.23 (21)    &    31.67(4) &  14.58 (21)  \\  
\hei\ 6678                &     11.92 ( 6) &     5.31 (22)    &    12.89(4) &   5.74 (22)  \\  
\sii\ 6716                &     11.90 ( 6) &     5.23 (22)    &     8.04(6) &   3.53 (22)  \\  
\sii\ 6731                &      9.39 (10) &     4.11 (24)    &     6.17(6) &   2.70 (23)  \\  
\hei\ 7065                &      5.25 (32) &     2.07 (40)    &     3.61(6) &   1.42 (25)  \\  
\ariii\ 7136              &     33.78 ( 6) &    13.04 (26)    &    27.83(4) &  10.75 (25)  \\  
\ariv+\heii\              &  	           &                  &     2.75(6) &   1.05 (26)  \\  
\oii\  7320               &      9.32 (14) &     3.42 (30)    &     9.15(7) &   3.36 (27)  \\ 
\oii\  7330               &      6.14 (14) &     2.25 (30)    &     6.43(7) &   2.35 (27)  \\ 
\heii\ 7592               &                &                  &     2.54(6) &   0.87 (29)  \\ 
\ariii\ 7751              &                &                  &     7.26(6) &   2.38 (30)  \\ 
\heii\ 8237               &                &                  &     6.28(6) &   1.85 (33)  \\ 
\hi\ 9015                 &                &                  &     6.73(9) &   1.72 (37)  \\ 
\siii\ 9069               &                &                  &    63.97(9) &  16.20 (37)  \\ 
\hline\\          
\end{tabular}
\newline $^\star$ Additional uncertainty as deblending from a faint
nearby line was required.
\label{T-linefluxes}
\end{table}

The strong \oiii\ and \heii\ lines in the GTC spectra confirm that
\n\ is indeed a planetary nebula of moderately high excitation. 
Its standard PN name is therefore PN~G086.9--03.4.
The nebular reddening was computed in the usual way from the hydrogen
Balmer decrement using the \ha, \hb, \hg, and \hd\ lines.  A weighted
average value of \cb=0.95$\pm$0.25 is obtained using the
reddening law of \cite{fitzpatrick04} and $R_\mathrm{V}$$=$3.1, and
adopting the electron density and temperature as determined below from
the \sii\ and \oiii\ lines, respectively.
This corresponds to $A_\mathrm{V}$=2.0$\pm$0.6~mag.  This \cb\ value
was adopted to deredden the fluxes in Tab.~\ref{T-linefluxes}: the
quoted errors on the dereddened line fluxes include the uncertainty on
the \cb\ value.  Physical and chemical conditions in the gas were then
computed with the same methodology as in \cite{c11}, using the {\sc
  python}-based {\sc PyNeb} package \citep{lms14}. They are listed in
Tab.~\ref{T-chem}. An electron density \Ne=125$\pm$40~cm$^{-3}$ is
obtained from the \sii~6731/6717 line ratio. An electron temperature
\Te=13050$\pm$850~K is determined from the \oiii~(5007+4959)/4363
ratio.  The lack of measurement of the electron temperature for
low-ionization stages, as well as substantial errors in the flux of
several important emission lines, prevent a complete chemical analysis
of the nebula, that is deferred to a future work.  However, very
interesting trends emerge from the analysis of the present
spectra. The significant error in the helium abundance in
Tab.~\ref{T-chem} reflects the dispersion in the \hei\ ionic
abundances estimated from different lines.  In particular, the ionic
abundance estimated for the adopted \Te\ from the (fainter) \hei~4471
and 7065 lines is significantly smaller than the values obtained from
the (stronger) \hei~5876 and 6678 lines. This discrepancy might be
relevant to understand the physical conditions in the nebula, as
discussed in Sect.~\ref{S-chem}.  For oxygen, nitrogen, and sulphur,
the quoted errors mainly reflect possible different choices for the
electron temperature of the low-excitation ions, taken either as
\Te\ computed from the \oiii\ lines (see above), or adopting
\Te=10000~K using the recipes in \cite{kb94} and \cite{m03}. Note also
that the O$^+$ abundance, and hence the N$^+$/O$^+$ ratio, are
determined using the \oii~7320,7330 doublet, that is quite sensitive
to the adopted \Te. Also, it tends to overestimate the O$^+$ abundance
compared to the \oii~3729,3729 doublet (see e.g. \citealt{m10}). It is
therefore possible that the total oxygen abundance is lower than
quoted in the table, and that $\log$(N/O) is higher.

In the blue spectrum, the \cii\ recombination line at
  4267~\AA\ line and a blend of lines around 4650~\AA\ are detected
  (Fig.~\ref{F-spec}). Most likely they originate in the nebula, as their
  emission is clearly spatially extended and do not obviously appear
  in the nebula--subtracted spectrum of the central star (see
  Sect.~3.2). The 4650 blend is probably composed of O, N, and C
  recombination lines, but spectral resolution is too low to attempt a
  detailed identification.  The strength of these
  recombination lines is unusual for such a low density nebula.  For the
  above range of \Te\ considered for low-ionization atoms, the flux of
  \cii4267 listed in Table~\ref{T-linefluxes} (but note the
  50\%\ error, due to its faintness plus the need to deblend from an
  even fainter redder line) would imply a remarkably high carbon
  abundance (Tab.~\ref{T-chem}). This is another potentially important
  datum (see Sect.~\ref{S-chem}).


\begin{table}
\caption{Physical conditions and chemical abundances by number 
in \n.}
\centering
\begin{tabular}{ll}
\hline\hline
\multicolumn{2}{l}{Physical conditions:}  \\[2pt]
\Ne(\sii)   & 125$\pm$40 cm$^{-3}$ \\
\Te(\oiii)  & 13050$\pm$850~K      \\
\hline
\multicolumn{2}{l}{Abundances:}  \\[2pt]
He/H                         & 0.18$\pm$0.04   \\
$\log$(O/H)+12               &  8.25$\pm$0.15    \\
$\log$(S/H)+12               &  6.25$\pm$0.10      \\
$\log$(C$^{++}$/H)+12$^{\dag}$ &  9.5$^{+0.2}_{-0.3}$             \\
$\log$(N$^+$/O$^+$)          & $-$1.3$\pm$0.2     \\
\hline\\[-7pt]
\end{tabular}
\newline
$^{\dag}$ From the uncertain measurement of the\\ 
\cii~4267 recombination line, see text\\
\label{T-chem}
\end{table}

The nebula is a radio source detected by the NRAO VLA Sky Survey
\citep{c98} with an integrated flux of 3.6$\pm$0.5~mJy at 1.4~GHz. It
was not resolved by the radio beam, implying a size smaller than
74$''$$\times$46$''$. An average extinction can be estimated comparing
the radio and \ha\ fluxes \citep[cf.][eq. IV-26]{p83}.  The observed
integrated \ha\ flux of the nebula was estimated from the IPHAS
images, using the DR2 photometric calibration \citep{bar14}. It is
computed to be F(\ha)=8.5\,~10$^{-13}$~erg~cm$^{-2}$~s$^{-1}$, after
removing the estimated contribution of the \nii\ doublet to the IPHAS
images.  Assuming that the radio emission is optically thin and that
all the extinction is external to the object, we obtain
\cb=1.23$\pm$0.30.  This value is in fair agreement with the value
deduced above from the Balmer decrement.

The GTC spectra lack the resolution needed to accurately measure
the systemic velocity of the nebula. Nevertheless, using bright
nebular lines relatively close to night-sky lines, we can conclude that
there is no indication that \n\ strongly deviates from the radial
velocities expected for the Galactic rotation curve in the direction
of the source. 

\begin{figure*}
\centering
\includegraphics[width=1.0\hsize]{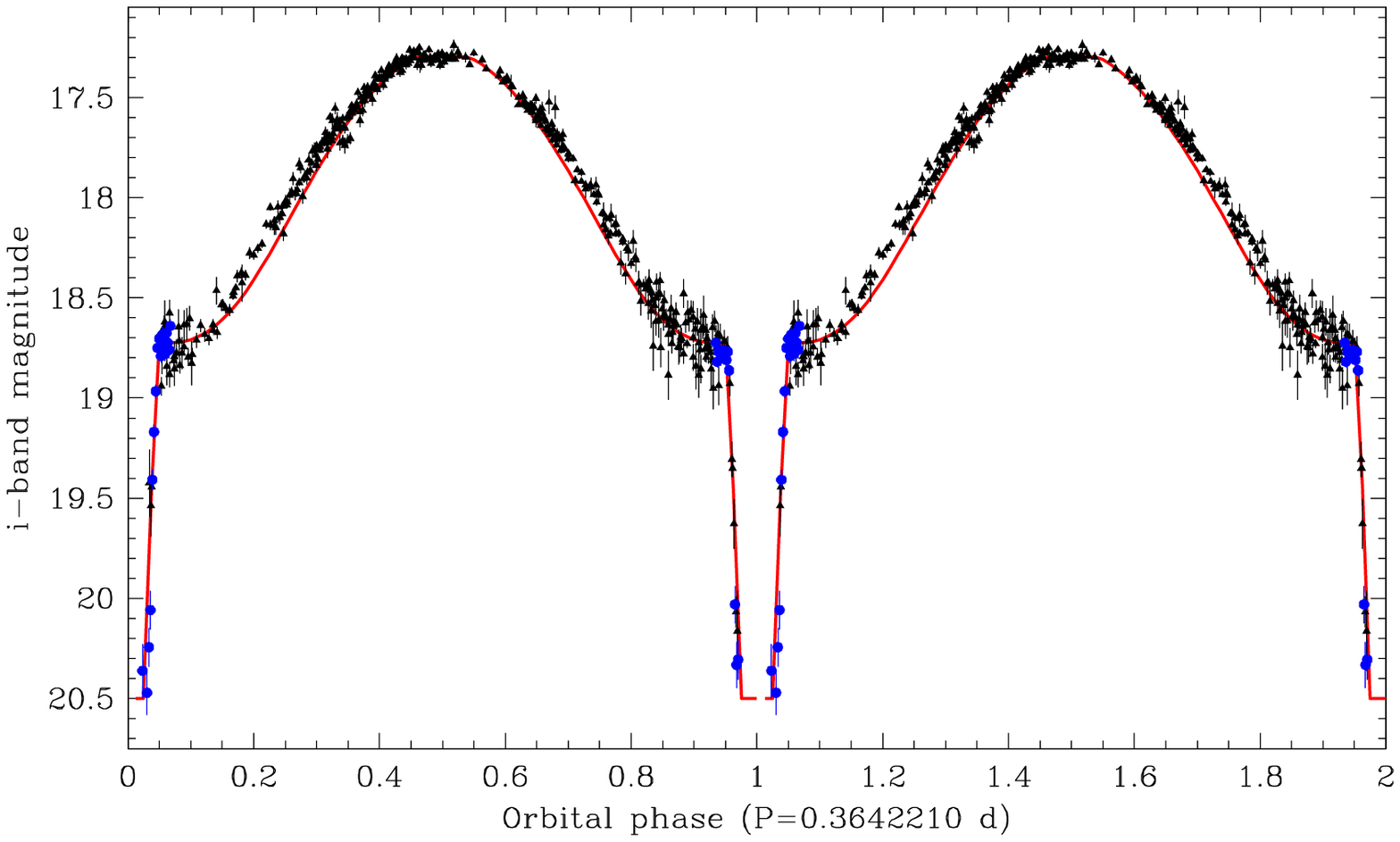}
\caption{SDSS $i$-band light curve of \cs\ from our own
    photometric monitoring, folded on the period of 8.74 hours
    according to the ephemeris given in the text. Magnitudes obtained
    with different telescopes have been scaled using field stars 
    to match the IPHAS photometry \citep{bar14}. The orbital
  cycle has been repeated twice for clarity.  No phase binning has
  been applied to data. (Black) triangles are measurements from all
  telescopes except for the 4.2m~WHT, which is represented by the
  (blue) full circles. The scatter close to light curve minimum is
  caused by the lower S/N ratio of the data from the smaller
  telescopes. The illustrative 
model presented in the text is indicated as a solid
  (red) line.}
\label{F-lightcurve}
\end{figure*}

\subsection{The central star(s)}

According to the IAU-approved IPHAS nomenclature, the central star
should be named \cs.  Its light curve in the SDSS $i$--band from
  our own photometric monitoring, folded on the period of 8 hours and
44 minutes determined below, is shown in Fig.~\ref{F-lightcurve}.

The light curve was determined as follows from observations obtained
in several nights and different telescopes (see Sect.~2). Note
that its determination is complicated by the presence of nearby stars,
and in particular of a relatively bright one located 3$''$.5 south of
\cs. To avoid its contamination, all measurements obtained with a
seeing worse than 1$''$.7 were rejected. Data obtained with the
smaller telescopes during eclipses were also not considered, because
of their insufficient S/N ratio.  The first photometric series,
obtained at the IAC80 telescope on October 31, showed an eclipse with
a duration of $\approx$50~min.
On the following night, the light curve displayed a maximum.  No
observations in the same night contained two consecutive eclipses, and
the length of the longest monitoring period with a recorded eclipse
pointed to an orbital period in excess of four hours. Both a sine fit
and an analysis-of-variance \citep[AOV,][]{sc96} periodogram of all
the photometric observations (with the eclipses masked out) suggested
a period of 8.74 hours.  The fact that the first two observed eclipses
(October 31 and November 23) were separated by an integer multiple of
8.74-h cycles, confirmed that this periodicity was the orbital period
of the binary system.  In order to compute an accurate linear
ephemeris we measured the times of all available light-curve minima
and eclipses (orbital phase 0), and maxima (orbital phase 0.5) by
Gaussian fitting. Details are given in Table~\ref{table_ephem}. The
resulting ephemeris is:
\begin{eqnarray}
\centering
 T_0(\mbox{HJD}) = 2456597.50034(55) + 0.3642210(44) \times E~,\nonumber
\end{eqnarray}
where $T_0$ is the time of inferior conjunction of the companion star
(i.e. the centre of the eclipse). The figures quoted in
parentheses are the formal statistical errors (given in units of the
last decimal place) of the linear fit to the timings shown in
Table~\ref{table_ephem}.

\begin{table}
\setlength{\tabcolsep}{0.95ex}
\caption[]{Timings of the observed centres of the eclipses (orbital
  phase 0) and light curve maxima (0.5).}
\label{table_ephem}
\begin{center}
\begin{tabular}{lc}
\hline\noalign{\smallskip}
~~~~~~~~~~$T_0$ & ~~~~~~~~~~~~~~~Cycle ($E$)\\
($\mathrm{HJD} - 2450000$) & ~~~~~~~~~~~~~~~\\  
\hline\noalign{\smallskip}
6597.51(1)    & ~~~~~~~~~~~~~~~0     \\
6598.4113(6)  & ~~~~~~~~~~~~~~~2.5   \\
6614.249(2)   & ~~~~~~~~~~~~~~~46    \\
6620.445(2)   & ~~~~~~~~~~~~~~~63    \\
6621.3569(7)  & ~~~~~~~~~~~~~~~65.5  \\
6622.449(2)   & ~~~~~~~~~~~~~~~68.5  \\
6643.3922(1)  & ~~~~~~~~~~~~~~~126   \\
\hline\noalign{\smallskip}
\end{tabular}
\end{center}
\end{table}

The light curve shows a deep eclipse ($\sim$1.6~mag) at the minimum of
an ample sinusoidal modulation ($\sim$1.5~mag, among the largest found
in PN central stars).  The sampling during the eclipse is
  incomplete because, on the night when the 4.2m~WHT was used, we
  switched to spectroscopic mode just after the end of the eclipse
  ingress in an attempt to obtain a 20--min spectrum of the
  non-irradiated side of the companion star. The signal on the
  spectrum was too faint to be useful, also because of the high
  background produced by the full moon in a night in which dust
  pollution was present in the atmosphere. However, the two
  photometric points closer to phase 1 suggest that the eclipse bottom
  was reached at an i-band magnitude between 20.4 and 20.5.  The
light curve of \cs\ is similar to that of Hen~2-11 \citep{j14}, but
with a shorter period and a more pronounced luminosity variation. As
in the latter, the interpretation is that an irradiated companion
produces the sinusoidal part of the curve during its orbital motion
\citep[see e.g.][]{c11}, and a (total) eclipse of the hot post-AGB
star occurs at inferior conjunction, when the companion faces us its
non--irradiated hemisphere.  If so, data imply that, at these red
wavelengths, the luminosity ratio between the post-AGB star and the
irradiated side of the companion is around one fourth at light curve
maximum. On the contrary, at the minimum of the sinusoidal modulation
(inferior conjunction), the post-AGB star would dominate the emission
over the non-irradiated side of the companion by a factor of three.
The luminosity of the irradiated side of the companion would be some
fifteen times larger than the non-irradiated hemisphere: this implies
that even at small orbital phases ($\ge$0.1) the irradiated hemisphere
dominates over the non-irradiated one, and therefore the latter can be
properly studied only during the main eclipses.  The shallow eclipse
at light curve maximum indicates that only a small fraction of the
projected surface of the companion is eclipsed by the compact post-AGB
star.

Further insights can be gained by modelling the light curve using the
{\sc
  nightfall}\footnote{\href{http://www.hs.uni-hamburg.de/DE/Ins/Per/Wichmann/Nightfall.html}
  {http://www.hs.uni-hamburg.de/DE/Ins/Per/Wichmann/Nightfall.html}}
code. All parameters were varied over a wide range of physical
solutions, with the best-fitting model being selected as the one with
the lowest $\chi$$^2$ fit to data.  Given the large irradiation effect
in the system, detailed reflection was employed in the modelling (with
3 iterations) in order to properly treat the irradiation of the
secondary by the primary. A model atmosphere was used for the lower
temperature (secondary) component with solar metallicity and
$\log$~g$=$4.5 \citep{k93}, and a blackbody for the primary. An
  illustrative model light curve is shown in Fig.~\ref{F-lightcurve}
  as a red solid line. The corresponding binary parameters (where
  index 1 refer to the post--AGB star, and 2 to the companion) are:
  T$_1$=85\,000~K, R$_1$=0.11~R$_\odot$, T$_2$=4\,000~K,
  R$_2$=0.65~R$_\odot$, $M_1/M_2$=0.7, and i=90\degr.  In
  general, a reasonable fit to the data is obtained, but the following
  cautionary points should be noted.  First, there is a small but
  systematic deviation around quadratures, where we observe some
  excess light compared to the model. This might be related to the
  complex heating/diffusion process in such a highly irradiated star,
  that might not be fully treated by {\sc nightfall}. Second, the fit
  of the eclipse is not entirely satisfactory, as the ingress and
  egress are ``slower'', and the depth possibly larger, than indicated
  by the observations.  If one tries to fit a shallower eclipse, then
  the model temperature of both stars significantly increases, and the
  mass ratio raises above unity towards unlikely large
  values. Therefore the binary parameters, and in particular the mass
  ratio, are loosely constrained by the present data, only limited
  to $i$--band light curve.  It is clear that additional observations,
  including photometry at other wavelengths, spectroscopy, and radial
  velocity studies, are needed to produce a robust model, and properly
  constrain the binary parameters.

The high orbital inclination from the fit, that is required in order
to have eclipses, is consistent with the apparent morphology of the
nebula if the axes of the orbit and the nebula coincide, as it has
been found in other PNe with close binary central stars
\citep[e.g.][and references therein]{j12,t12,h13}.

According to the ephemeris above, the blue GTC spectrum of \n\ was
taken at orbital phase between 0.06 and 0.09, i.e. right after the end
of the eclipse of the hot post-AGB star.  On the contrary, the red GTC
spectrum was taken at phase between 0.87 and 0.90, i.e. symmetrically
to the blue spectrum and right before the eclipse. At these
phases, no radial velocity shifts due to the orbital motion can be
detected at the low resolution of our GTC spectra.  In these
orbital configurations, following the discussion above the emission
from the stellar system would be mainly due to the hot post-AGB star,
with a smaller contribution of the companion.  The observed stellar
spectrum, after nebular subtraction,
is relatively flat and basically featureless.  No absorption lines of
\hi, \hei\ or \heii\ from the ionizing post-AGB star are detected, but
they are strongly affected by the subtraction of the corresponding
nebular emission lines at the limited spectral resolution of the
present observations.  The typical emission lines from the irradiated
hemisphere of the companion such as the 4650~\AA\ blend \citep[see
  e.g.][]{c11} are not seen either. And, finally, no obvious stellar
feature from the cooler side of the companion is detected.  New
observations, at higher spectral resolution and at various orbital
phases, are needed to directly investigate the central stars.

\subsection{Distance, mass, and spectral energy distribution}

The available data only allow a rough distance determination for \n.
A first estimate can be done using the relation between the H$\alpha$
surface brightness and radius of the nebula by \cite{f06}.  A distance
of 5$\pm$1~kpc is found by adopting the relation suitable for PNe with
close binary central stars.  A similar result, 5.6~kpc, is obtained
using the Shklovsky/Daub statistical method as calibrated by
\cite{s08}.  

An alternative distance estimate can be obtained from the spectral
type of the companion, if we assume the (uncertain) stellar
  temperature derived from modelling the light curve. This would
  indicate a late--K spectral type for the secondary \citep{bo13}. From
  the light curve, we find that the apparent luminosity of the
  night-side hemisphere of the companion is i=20.4~mag, or i=19.1
    after dereddening.  Using the intrinsic colours and absolute
  magnitude of K dwarfs from \cite{bil09},
we obtain a distance of 2.5~kpc.

For a distance $\le$5 kpc, and adopting a standard value of 0.4 for
the filling factor, the ionized mass of the nebula is then estimated to be
$\le$10$^{-1}$ ~M$_\odot$.

\begin{figure}
\centering
\includegraphics[width=1.0\columnwidth]{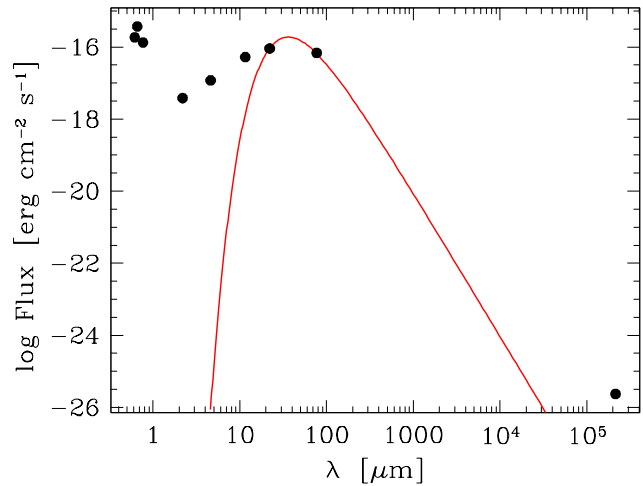}
\caption{Spectral energy distribution of \n. Points are observational
  data (their errors are smaller than the symbols),
while the (red) solid line is a blackbody with a temperature of 80~K 
(see text).}
\label{F-sed}
\end{figure}

Fig.~\ref{F-sed} shows the Spectral Energy Distribution (SED) of the
object. The IPHAS and UKIDDS-K data (0.6 to 2.1~$\mu$m) correspond to
the central star(s) whereas all other IR data (WISE W2, W3, W4, and
the AKARI 90~$\mu$m flux) correspond to the extended nebula. Note that
2MASS J, H and Ks, and WISE W1 fluxes are not included in the figure
because they mostly originate from the field star 3$''$.5 SW of \cs.
The 1.4~GHz radio flux from \cite{c98} is also plotted. The SED in the
IR region is very broad, indicating a range of dust temperatures, and
it shows a maximum around 70~$\mu$m; both are normal characteristics
of PNe \citep[c.f.][]{a12}. Indeed, the WISE W2--W4 and W3--W4 colours
are typical of PNe, and very dissimilar to those of young stellar
objects and symbiotic stars, confirming the nature of \n\ as a PN. The
temperature of the coldest detected dust grains can be estimated from
the 22 and 90~$\mu$m fluxes and is around 80~K (solid line in the
figure). The measured total flux, integrating below the SED from 0.6
to 90~$\mu$m, is 5.5~10$^{-11}$~erg~cm$^{-2}$~s$^{-1}$ and it
increases to 1.3~10$^{-10}$~erg~cm$^{-2}$~s$^{-1}$ if the expected
fluxes from a 80~K blackbody longward of 90~$\mu$m are included. For a
distance of 5~kpc, the latter value would imply a total luminosity
L$\ge$100~\lsun, the limit coming from the fact that a substantial
population of colder dust grains, with a large FIR-submm contribution
to the SED, are generally expected in PNe.

\section{Discussion}


The short orbital period of \cs\ indicates that, during the red-giant
phase of the star that has produced the PN, the system went through a
common-envelope (CE) phase that resulted in the shrinkage of the orbit
to the current size \citep{p76,il93}. The CE phase is generally
thought to end with a stellar merger or the complete ejection of the
stellar envelope, but the process and its timescales are utterly
unknown \citep{i13}.

Post-CE PNe are key objects to better understand CE evolution. In
particular, the two most relevant aspects of the present observations
are: {\it i)} the nebular morphology, which provides information about
the history and geometry of the CE ejection process, and {\it ii)} the
nebular chemistry, which in principle allows discussion of the whole
evolution of the systems, because it depends on the original chemical
content of the stellar progenitors, on the chemical enrichment during
stellar evolution, and on the effects of the binary interactions.

\subsection{CE ejection}

Concerning morphology, the message given by \n\ is a clear one: its
multiple pairs of lobes seem to indicate that the (common?) envelope
was ejected in at least two distinct episodes.  Similar cases are
M~2-19 and Abell~41 \citep{m09b} -- although for the latter no
kinematical identification of distinct outflows has been obtained
\citep{j10} -- and perhaps ETHOS1 \citep{m11}.  In other post-CE PNe
there is evidence of pre-CE mass loss, mainly in the form of light
jets likely ejected by accreting companions, the Necklace nebula being
the most convincing case \citep{c11,m13,t14}. But these jets are secondary
morphological components as most of the mass is in the inner bodies of
the nebulae.  On the contrary, in \n\ two apparently distinct, major
bipolar outflows are found.  If this is confirmed by the kinematical
analysis of the nebula, that is needed to determine the ages of each
bipolar outflow, \n\ would add puzzling information about the mass
loss processes during CE evolution, opening the possibility that the
CE is ejected in distinct episodes. Alternatively, the outer pair of
lobes could have been produced before the CE phase, for instance as an
ablation flow left behind by a jet that has now vanished.  Indeed, the
truncated outer lobes of \n\ are similar to those observed in several
pre-PNe, whose formation is often ascribed to the action of jets
\citep{b14}.  Another possibility is that the two pairs of lobes are
practically coeval, as in the case of the symbiotic nebula Hen~2--104
\citep{c01}, a result that is however difficult to explain
theoretically. The additional possibility that the apparent double
pair of lobes in \n\ are just the result of an excitation gradient, as
suggested in the case of M~2--9 \citep{s05}, is instead less
likely. In any case, the mild bipolar morphology of \n\ and similar
post-CE PNe can be produced by fast winds (either isotropic or
collimated) impinging on flattened mass distributions with modest
equatorial--to--polar density contrasts \citep{i89}. Such a geometry
is expected in the CE ejection process \citep{s98}, or can be easily
produced right before it during a period of stable Roche-lobe mass
transfer onto the companion, or by gravitational focussing of the AGB
progenitor when the binary system was much wider \citep{mm99}.

\subsection{Hints on the chemistry of post--CE nebulae}
\label{S-chem}

\begin{figure}
\centering
\includegraphics[width=1.0\columnwidth]{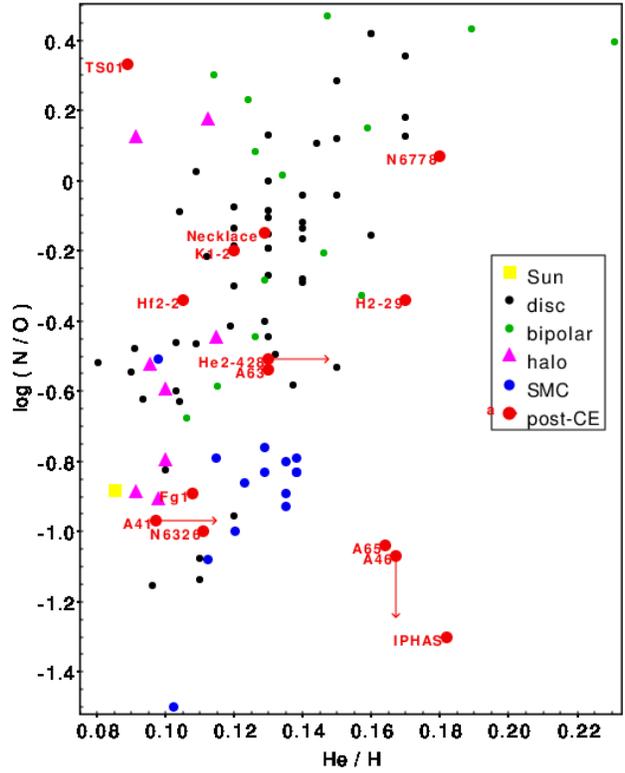}
\caption{Chemical abundance ratios for post--CE PNe and other
  representative samples of PNe.  \n\ is indicated by the label
  ``IPHAS''. Abundance limits are indicated by arrows.}
\label{F-chemistry}
\end{figure}

Standard analysis of the GTC spectra seems to indicate that \n\ has
peculiarly low sulphur and nitrogen abundances, but a high helium (and
perhaps carbon) content.  In general, information about the chemical
abundances of post--CE PNe is sparse, and in several cases quite
uncertain because many of these nebulae have a low surface brightness.
Fig.~\ref{F-chemistry} shows the relation of two abundance ratios that
do not suffer from large uncertainties as other elemental abundances:
N/O, commonly assumed to be equal to the observed N$^+$/O$^+$,
vs. He/H.  They are indicated for a sample of fourteen post-CE PNe
taken from the articles of
\cite{kb94,pb97,e03,l06,r01,p04,g09,st10,m10,c11}; and this study for
\n.
These post--CE nebulae are indicated in the figure by full (red)
circles.  Their abundances are compared to representative samples of
PNe, as indicated in the legenda of the figure: ``classical
bipolars''\footnote{See \cite{cs95} for their definition.}  from
\cite{pc98}; Galactic disc PNe from the works of \cite{h10} and
\cite{m10}; Galactic halo PNe from \cite{h97}; and SMC PNe from
\cite{s10}.

The figure suggests that a considerable fraction of post--CE PNe have
distinctive properties compared to the bulk of Galactic PNe. TS~01 is
a Galactic halo PN with a double--degenerate binary central star which
holds the record of being the most oxygen--deficient PN \citep{st10},
where part of its depletion being likely related to the binary
evolution. 
About half of the other thirteen post--CE nebulae have low N/O
ratios\footnote{This also applies to the other post-CE PNe M~2--19 and
  M~3--19, which are not included in the figure because their He/H
  abundance is uncertain \citep{cui00}.} compared to the bulk of
Galactic disc PNe and, in spite of the fact that many of them have a
highly collimated morphology, lower than classical bipolars. A prime
example is M~2--19 that has a bipolar PN and a post-CE central star
\citep{m08}.  On the contrary, their He/H abundance is in the range of
average PNe, and in some of them is remarkably high, He/H$>$0.16. The
increase in N/O that accompanies the He/H overabundances in classical
bipolars is therefore not observed in most post-CE PNe. Their lowest
N/O ratios are only comparable to that of the SMC PNe. In this
respect, the low sulphur content of \n\ is also found in several other
post-CE PNe \citep{kb94,pb97,r01,e03}, and again is within the range
displayed by the low-metallicity SMC and Galactic halo PNe.


Carbon deserves an additional comment. Even if obtained from an
uncertain flux measurement, the large C$^{++}$ abundance of \n\ calls
the attention because similar results were obtained for the other
binary PNe A~46 and A~63 \citep{pb97}, and Hf~2--2 \citep{l06}.  If
the abundance from the \cii~4267 recombination line is compared to the one
computed from the \oiii\ collisionally excited lines, we derive
C$^{++}$/O$^{++}$=35, which transforms into a C/O ratio between 23 and
32 using the {\it icf} values by \cite{d14}.  Such a ratio is larger
by almost one order of magnitude than expected in any carbon-rich
(single) AGB precursor.  However, it is well known that in
photoionized nebulae optical recombination lines (ORLs) provide
abundance values that are systematically larger than those obtained
using collisionally excited lines (CELs). The so-called abundance
discrepancy factor ({\it adf}) is usually between 1.5 and 3 \citep[see
  e.g.][]{g07,l12}, but in PNe it has a significant tail extending to
much larger values.  In particular, it is striking that the Galactic
PN with the largest known {\it adf}, around 70, is the aforementioned
post--CE PN Hf~2--2 \citep{l06}. In that nebula, recombination lines,
including those of H and He, mainly come from a cold (\Te$\le$1000~K),
high metallicity nebular component whose mass in metals, in that
specific object, is comparable to that of the hotter gas where CELs
are produced.

The case of Hf~2--2 suggests that a similar physical situation (and
comparably large {\it adf} values) might also apply to \n\ and the
other post--CE nebulae that display anomalously high C$^{++}$
abundances from the \cii~4267 line. Existing spectra do not allow to
thoroughly test this hypothesis, as no O recombination lines can be
safely identified and used\footnote{The
  4645~\AA\ blend (Fig.~\ref{F-spec}) is likely to be a mixture of O,
  N, and perhaps C lines, and significant NLTE effects could
  occur.}. However, in the case of \n\ \citep[but not A~46 and
  A~63,][]{pb97}, in addition to its large C$^{++}$ abundance, the
discrepancy between the He$^+$ ionic abundances from different lines
obtained adopting \Te(\oiii) would also support the existence of a
cold ionized nebular component. Indeed, if the \hei~5875/4471 and
\hei~6678/4471 line ratios are inserted in the \Te\ diagnostic diagram
of Figure~4 of \cite{l06}, \Te$\approx$10$^3$ is estimated for the
\hei\ emitting region (but note that the measurement of the flux of
\hei~4471 has a non-negligible error).  If such a double nebular
component is present in \n\ as well, the determination of the
reddening and of all elemental abundances would be severely affected.
The sub-solar abundances computed from CELs for \n\ may then be a
spurious result, as it was shown for oxygen in Hf~2--2 \citep{l06}.


Summarising, even if existing data are of limited quality, there is
tempting indication that a significant fraction of post--CE PNe have a
peculiar chemistry. They might point to low-metallicity progenitors,
perhaps with some additional peculiarities (such as a sometimes
remarkably high He content) related to the binary evolution. However,
the well-studied case of Hf~2--2 \citep{l06}, and the preliminary
analysis of \n\ and a couple of other PNe, suggests that a correct
chemical analysis requires a thorough investigation of abundances from
ORLs and a deep look at the abundance discrepancy problem in these
objects.

\subsection{Nebular mass}

A related, relevant property might be the very small ionized mass --
between 10$^{-3}$ and 10$^{-1}$~\msun\ -- estimated in the post-CE
nebulae A~63 and A~46 \citep{b94}, Hf~2--2 \citep{l06}, and \n\ (all
them, with the apparently large C$^{++}$ abundances discussed above).
This is also supported by the systematically lower luminosity of
  post--CE nebulae in the surface brightness-radius relation of
  \cite{f08}, compared to the whole sample of Galactic PNe. This
would imply that the stellar envelope in the CE phase is very small,
either because of a low initial stellar mass, or because much of the
envelope was lost or transferred to the companion before the CE event,
during a long-lasting, stable mass-transfer phase.  The latter
hypothesis seems to apply to another post--CE PN, the Necklace nebula,
whose stellar progenitor is supposed to be a carbon AGB star
(therefore relatively massive, $>$2~\msun, \citealt{c11,m13}), but the
ionized mass of the nebula is estimated to be only 0.1~\msun, and
there is evidence of conspicuous pre-CE mass transfer onto the
companion.  A small nebular mass may also facilitate the detection of
(also small) high metallicity components that are often invoked to be
at the origin of the abundance discrepancy.

\section{Summary and perspectives}
\label{S-summary}

We have presented nebular spectroscopy and the light curve of the
  new post-CE PN \n.

The nebula has a remarkable morphology, suggesting the occurrence of
multiple ejection events in the CE phase.  As mentioned above, the key
to recover the mass loss history constraining the CE (and pre-CE)
evolution, is to study the nebular dynamics via spatially resolved
high-resolution spectroscopy.

The chemical abundances of \n\ and other similar PNe are also very
interesting. Together with the evidence for low ionized masses, the
observed abundances seem to point to low-mass progenitors for a
significant fraction of post--CE nebulae. In addition, we suggest that
several post-CE PNe may have large {\it adf} values, as shown for
Hf~2--2 by \cite{l06}. One possible explanation is that accretion onto
a companion lightens the common envelope, producing a low mass PN
where the ORL emitting regions are more readily revealed.

The potential of these results calls for additional work, both
observational and theoretical.  Detailed modelling of the emission
nebulae is needed to determine the gas mass involved at any stage
around the CE phase. This methodology has not been applied
  systematically to this kind of PNe. To reach this goal,
model-independent distance determinations (as provided for instance by
extinction distances, or obtained from modelling of the stellar
atmospheres) are also required.  The chemistry of \n\ and other post-CE PNe
also needs to be better determined, with deep spectroscopic
observations and detailed photoionization modelling. The increasing
evidence of a peculiar nebular chemistry and the possible link with
the abundance discrepancy problem make this research line particularly
attractive.

Finally, the eclipsing nature of \cs\ makes it a favourable target to
determine the stellar parameters via spectroscopy at specific
orbital phases, and in particular the stellar masses via radial
velocity studies. With the global information of the binary system,
one could then try to reconstruct its entire evolution.

\section*{Acknowledgments}

Based on observations obtained with: the 10.4m Gran Telescopio
Canarias (GTC), the 4.2m~WHT and the 2.5m~INT telescopes of the Isaac
Newton Group of Telescopes, the 2.6m Nordic Optical Telescope operated
by NOTSA, and the 0.8m~IAC80 telescope, operating on the islands of La
Palma and Tenerife at the Spanish Observatories of the Roque de Los
Muchachos and Teide of the Instituto de Astrof\'\i sica de
Canarias. Also based on observations with the 0.6m telescope at Tartu
Observatory (Estonia). Some of the data were obtained with ALFOSC,
which is provided by the Instituto de Astrof\'\i sica de Andaluc\'\i a
under a joint agreement with the University of Copenhagen and NOTSA.
We are grateful to the time allocation committee (CAT) for awarding us
Director Discretionary Time at the WHT on December 16, 2013.  

We thank Pedro A. Gonz\'alez Morales, Cristina Zurita and Olga Zamora
for taking some of the photometric observations. We also thank Geert
Barentsen, Janet Drew, and Mike Barlow for their help in determining
the IPHAS photometry of the nebula. Finally, we thank the referee,
Orsola De Marco, for her suggestions to improve the content of the
article.

RLMC and AM acknowledge funding from the Spanish AYA2012-35330 grant.
PRG is supported by a Ram\'on y Cajal fellowship (RYC--2010--05762),
and acknowledges support provided by the Spanish AYA2012--38700 grant.
TE and TL acknowledge support by a targeted financing project
SF00600399S08 of the Estonian Ministry of Education and Research.  


\label{lastpage}


\begin{thebibliography}{100}
\bibitem[{Acker} \& {Le D\^u}(2014)]{acker14}
{Acker} A., Le D\^u P., 2014, L'Astronomie, n. 68, p. 40 
\bibitem[{Acker} {et~al.}(1992)]{acker92}
{Acker} A., {Marcout} J., {Ochsenbein} F., Stenholm B., Tylenda R., Schohn C., 
1992, {The Strasbourg-ESO Catalogue of Galactic Planetary Nebulae. Parts I, II}
\bibitem[{{Acker} {et~al.}(2012)}]{acker12}
{Acker} A., {Boffin} H.M.J., {Outters} N., Miszalski B., Sabin L.,
Le Du P., Alves F., 2012, Rev. Mexicana Astron. Astrof. 48, 223
\bibitem[{Anderson} {et~al.}(2010)]{a12}
{Anderson} L.D., Zavagno A., Barlow M.J., 
Garc\'\i a-Lario P., Noriega-Crespo A., 2012, A\&A, 537, A1 
\bibitem[{{Balick} {et~al.}(2014)}]{b14}
{Balick} B. et al. 2014, ApJ, in preparation
\bibitem[{{Barentsen} {et al.}(2014)}]{bar14}
{Barentsen}, G. et al., 2014, MNRAS, in preparation
\bibitem[{{Bell, Pollacco \& Hilditch}(1994)}]{b94}
Bell S.A., Pollacco D.L., Hilditch R.W., 1994, MNRAS, 270, 449
\bibitem[{{Bilir} {et~al.}(2009)}]{bil09}
Bilir S., Karaali S.Ak., Coskunoglu K.B., Yaz E., Cabrera--Lavers A., 
2009, MNRAS, 396, 1589
\bibitem[{{Bond}(2000)}]{b00}
Bond H.E, 2000, in Asymmetrical Planetary Nebulae III: from origins
to microstructures'', ASP Conf. Ser., Vol. 199, p. 115
\bibitem[{{Boyajan} {et~al.}(2013)}]{bo13}
Boyajian T.S. et al.,  
2013, ApJ, 771, 40
\bibitem[{{Condon} {et al.}(1998)}]{c98}
Condon J.J., Cotton W.D., Greisen E.W., Yin Q.F., Perley R.A., 
Taylor G.B., Broderick, J.J., 1998, AJ, 115, 1693
\bibitem[{{Corradi} \& {Schwarz}(1995)}]{cs95}
{Corradi} R.L.M., Schwarz H.E., 1995, A\&A, 293, 871
\bibitem[{{Corradi} {et~al.}(2001)}]{c01}
{Corradi} R.L.M., Livio M., Balick B., Munari U., Schwarz H.E., 
2001, ApJ, 553, 211
\bibitem[{{Corradi} {et~al.}(2011)}]{c11}
{Corradi} R.L.M. et al., 2011, MNRAS, 410, 1349
\bibitem[{{Cuisinier} {et~al.}(2000)}]{cui00}
Cuisinier F., Maciel W.J., K\"oppen J., Acker A., Stenholm B., 
2000, A\&A, 353, 543 
\bibitem[{Delgado--Inglada, Morisset \& Stasi\'nska}(2014)]{d14}
Delgado--Inglada G., Morisset C., Stasi\'nska G., 2014, MNRAS, submitted
\bibitem[{De Marco} {et~al.}(2007)]{dm07}
De Marco O., Wortel S., Bond H.E., Harmer D., 2007, in ``Asymmetrical
Planetary Nebulae IV'', published online at
http://www.iac.es/proyect/apn4, \#75
\bibitem[{De Marco} {et~al.}(2013)]{dm13}
De Marco O., Passy J.-C., Frew D.J., Moe M., Jacoby G.H., 2013, MNRAS, 428, 2118 
\bibitem[{{Drew} {et~al.}(2005)}]{d05}
{Drew} J. et al., 2005, MNRAS, 362, 753
\bibitem[{{Exter, Pollacco \& Bell}(2003)}]{e03}
Exter K.M., Pollacco D.L., Bell S.A., 2003, MNRAS, 341,1349
\bibitem[{{Fitzpatrick}(2004)}]{fitzpatrick04}
{Fitzpatrick} E.~L., 2004, in ASP Conf. Ser., Vol. 309, 
Astrophysics of Dust, A.N. {Witt}, G.C. {Clayton} \& B.T. {Draine} eds., p. 33
\bibitem[{Frew}(2008)]{f08}
Frew D.J., 2008, PhD Thesis, Macquarie University, Sydney, Australia
\bibitem[{Frew} \& {Parker}(2006)]{f06}
Frew D.J., Parker Q.A., 2006, in IAU Symp. 234, Planetary
Nebulae in our Galaxy and Beyond, M.J. Barlow \& R.H. Mendez eds., p. 49
\bibitem[{Garc\'\i a--Rojas} \& {Esteban}(2007)]{g07}
Garc\'\i a--Rojas J., Esteban C., 2007, ApJ, 670, 457
\bibitem[{G\'orny et al.}(2009)]{g09}
G\'orny S.K., Chiappini C., Stasi\'nska G., Cuisinier F., 2009, A\&A, 500, 1089
\bibitem[{{Henry} {et al.}(2010)}]{h10}
Henry R.C.B., Kwitter K.B., Jaskot A.E., Balick B., Morrison M.A., 
Milingo J.B.,  2010, ApJ, 724, 748
\bibitem[{{Howard, Henry \& McCartney}(1997)}]{h97}
Howard J.W., Henry R.C.B., McCartney S., 1997, MNRAS, 284, 465 
\bibitem[{{Huckvale}{ et al.}(2013)}]{h13}
Huckvale L. et al., 2013, MNRAS, 434, 1505
\bibitem[{Iben \& Livio}(1993)]{il93}
Iben I.Jr., Livio M., 1993, PASP, 105, 1373
\bibitem[{Icke, Preston \& Balick}(1989)]{i89}
Icke V., Preston H.L., Balick B., 1989, AJ, 97, 462
\bibitem[{Ivanova} {et~al.}(2013)]{i13}
Ivanova N., et al., 2013, A\&ARv, 21, 59
\bibitem[{Jones} {et al.}(2010)]{j10} 
{Jones} D. et al., 2010, MNRAS, 408, 2312
\bibitem[{Jones} {et al.}(2012)]{j12} 
{Jones} D., Mitchell D.L., Lloyd M., Pollacco D., 
O'Brien T.J., Meaburn J., Vaytet N.M.H., 2012, MNRAS, 420, 2271
\bibitem[{Jones} {et al.}(2014)]{j14} 
{Jones} D., Boffin H.M.J., Miszalski B., Wesson R., Corradi R.L.M., 
Tyndall A.A., 2014, A\&A, 562, A89
\bibitem[{{Kingsburg} \& {Barlow}(1994)}]{kb94} 
Kingsburgh R.L., Barlow M.J., 1994 MNRAS, 271, 257
\bibitem[{{Kurucz}(1993)}]{k93} 
Kurucz R.L., 1993, VizieR Online Data Catalog, 6039, 0
\bibitem[{{Liu}(2012)}]{l12} 
Liu X.W., 2012, in IAU Symp. 283, Planetary Nebulae: An Eye to the Future, 
Manchado, A. Stanghellini, L. \& Sch\"onberner, D. eds., p. 131
\bibitem[{{Liu} {et al.}(2006)}]{l06}
Liu X.W., Barlow M.J., Zhang Y., Bastin R.J., Storey P.J., 
2006, MNRAS, 368, 1959
\bibitem[{{Luridiana, Morissett \& Shaw}(2014)}]{lms14}
{Luridiana} V., Morissett C., Shaw R.A., 2014, A\&A, in preparation
\bibitem[{Magrini} {et al.}(2003)]{m03} 
{Magrini} L., Perinotto M., Corradi R.L.M., Mampaso A., 2003, A\&A, 400, 511
\bibitem[{{Mastrodemos} \& {Morris}(1999)}]{mm99}
{Mastrodemos} N., Morris M., 1999, ApJ, 523, 357
\bibitem[{{Milingo} {et al.}(2010)}]{m10}
Milingo J.B., Kwitter K.B., Henry R.C.B., Souza S.P., 2010, ApJ,  711, 619
\bibitem[{{Miszalski} {et al.}(2008)}]{m08}
{Miszalski} B., Acker A., Moffat A.F.J., Parker Q.A., Udalski A., 
2008, A\&A, 488, L79
\bibitem[{{Miszalski} {et al.}(2009a)}]{m09a}
{Miszalski} B., Acker A., Moffat A.F.J., Parker Q.A., Udalski A., 
2009a, A\&A, 496, 813
\bibitem[{{Miszalski} {et al.}(2009b)}]{m09b}
{Miszalski} B., Acker A., Parker Q.A., Moffat A.F.J., 2009b, A\&A, 505, 249
\bibitem[{{Miszalski} {et al.}(2011)}]{m11}
{Miszalski} B., Corradi R.L.M., Boffin H.M.J., Jones D., Sabin L.,
Santander-Garc\'\i a M., Rodr\'\i guez-Gil P., Rubio-D\'\i ez M.M., 
2011, MNRAS, 413, 1264
\bibitem[{{Miszalski, Boffin \& Corradi}(2013)}]{m13}
{Miszalski} B., Boffin H.M.J., Corradi R.L.M., 2013, MNRAS, 428, L39
\bibitem[{{Morris}(1987)}]{m87}
Morris M., 1987, PASP, 99, 1115
\bibitem[{{Nordhaus \& Blackman}(2006)}]{n06}
Nordhaus J., Blackman E.G., 2006, MNRAS, 370, 2004
\bibitem[{{Oke}(1990)}]{oke90}
{Oke} J.~B., 1990, AJ, 99, 1621
\bibitem[{{Paczynski}(1976)}]{p76}
Paczynski B., 1976, in IAU Symp. No. 73, Structure and Evolution of
Close Binary Systems, P. Eggleton, S. Mitton \& J. Whelan eds.,  p.75
\bibitem[{{Perinotto} \& {Corradi}(1998)}]{pc98}
Perinotto M., Corradi R.L.M., 1998, A\&A, 332, 721
\bibitem[{{Perinotto, Morbidelli \& Scatarzi}(2004)}]{p04}
Perinotto M., Morbidelli L., Scatarzi A., 2004, MNRAS, 349, 793
\bibitem[{{Pollacco} \& {Bell}(1997)}]{pb97}
Pollacco D.L., Bell S.A., 1997, MNRAS, 284, 32
\bibitem[{{Pottasch}(1983)}]{p83}
Pottasch S., 1984, In Planetary Nebulae, ISBN 90-277-1672-2, 
Astrophysics \& Space Science Library, vol. 107
\bibitem[{{Rodr\'\i guez, Corradi, \& Mampaso}(2001)}]{r01}
Rodr\'\i guez M., Corradi R.L.M., Mampaso A., 2001, A\&A, 377, 1042
%
\bibitem[{Sandquist} {et al.}(1998)]{s98} 
Sandquist E.L., Taam R.E., Chen X., Bodenheimer P., Burkert A., 
1998, ApJ, 500, 909
\bibitem[{{Santander-Garc{\'{\i}}a} {et~al.}(2013)}]{s14}
{Santander-Garc{\'{\i}}a} M., et al., 2014, A\&A, in preparation
\bibitem[{{Schwarz, Corradi \& Melnick}(1992)}]{scm92}
{Schwarz} H.E., {Corradi} R.L.M., Melnick J., 1992, A\&AS, 96, 23
\bibitem[{Schwarzenberg-Czerny(1996)}]{sc96}
Schwarzenberg--Czerny A., 1996, ApJ, 460, L107
\bibitem[{{Shaw} {et~al.}(2010)}]{s10}
Shaw R.A. et al., 2010, ApJ, 717, 562
\bibitem[{{Smith, Balick \& Gehrz}(2005)}]{s05}
Smith N., Balick B., Gehrz R.D., 2005, AJ, 130, 853 
\bibitem[{{Soker}(1997)}]{s97}
Soker N., 1997, ApJS, 112, 487
\bibitem[{Stanghellini, Shaw \& Villaver}(2008)]{s08}
Stanghellini L., Shaw R.A., Villaver E., 2008, ApJ, 689, 194
\bibitem[{{Stanghellini} {et~al.}(2009)}]{stan09}
Stanghellini L.,  Lee T.--H., Shaw R.A., Balick B., Villaver E., 
2009, ApJ, 702, 733
\bibitem[{{Stasi\'nska} {et~al.}(2010)}]{st10}
Stasi\'nska G. et al., 2010, A\&A, 511, A44
\bibitem[{{Tyndall}{ et al.}(2012)}]{t12}
Tyndall A.A., Jones D., Lloyd M., O'Brien T.J., Pollacco D., 
2012, MNRAS, 422, 1804
\bibitem[{{Tocknell, De Marco \& Wardle}(2014)}]{t14}
Tocknell J., De Marco O., Wardle M., 2014, MNRAS, 439, 2014
\bibitem[{{Vassiliadis} \& {Wood}(1994)}]{vawo94}
Vassiliadis E., Wood P.R, 1994, ApJS, 92, 125
\bibitem[{{Van Wickel} {et al.}(2014)}]{vw14}
Van Wickel H., Jorissen A., Exter K., Raskin G., Prins S., Perez-Padilla J, 
Merges F., Pessemier W., 2014, A\&A, preprint (arXiv:1403.2605)
\bibitem[{{Zuckerman, Becklin \& McLean}(1991)}]{z91}
Zuckerman B., Becklin E.E., McLean, I.S., 1991, ASP Conf. Ser.,
Vol. 14, p. 161
%
\end{thebibliography}
\end{document}